\documentclass[12pt]{article}

\usepackage{newtxtext,newtxmath}
\usepackage{graphicx}
\usepackage[letterpaper,margin=1in]{geometry}
\usepackage{longtable}

\renewenvironment{abstract}
	{\quotation}
	{\endquotation}

\date{}

\makeatletter
\renewcommand{\fnum@figure}{\textbf{Figure \thefigure}}
\renewcommand{\fnum@table}{\textbf{Table \thetable}}
\makeatother

\usepackage{scicite}
\usepackage{url}

\def\scititle{
   Volumetric nanoscale localization using engineered point spread functions in light sheet microscopy

}
\title{\bfseries \boldmath \scititle}

\author{
	R.~E.~Bautista~Gonzalez$^{1}$,
	R.~Mouthaan$^{1}$,
	A.~Upadhya$^{1,2}$,\and
    D.~J.~X.~Chow$^{1,2}$,
    K.~R. Dunning$^{1,2}$,
    K.~Dholakia$^{1,3\ast}$ \and
	\small$^{1}$Centre of Light for Life and School of Biological Sciences, Adelaide University, SA 5005, Australia.\and
	\small$^{2}$Robinson Research Institute, School of Pharmacy and Biomedical Science, Adelaide University, SA 5005, Australia.\and
    \small$^{3}$SUPA, School of Physics and Astronomy, University of St Andrews, North Haugh, Fife, KY16 9SS, UK.\and
	\small$^\ast$Corresponding author. Email: kishan.dholakia@adelaide.edu.au
}

\begin{document} 

\maketitle

\begin{abstract} \bfseries \boldmath

Nanoscale three-dimensional localization across large biological volumes remains an outstanding challenge in optical microscopy, with existing approaches typically limited by imaging speed, volumetric field of view and localization precision when required simultaneously. Here, we overcome these limitations by combining a twin Airy engineered point spread function with two-photon light sheet fluorescence microscopy, enabling nanoscale localization throughout large volumetric fields of view. Our framework explicitly incorporates the broadband fluorescence emission characteristic of biological fluorophores, ensuring accurate localization under realistic imaging conditions. We achieve localization precisions of $<20$~nm laterally and 42~nm axially over volumes measuring 295~$\mu$m  × 330~$\mu$m  × 100~$\mu$m , with a projected path to sub-10-nm localization in millimeter-scale specimens. Experiments in fluorescent bead phantoms and live mammalian oocytes confirm robust performance in both controlled and biologically complex environments. These results establish a scalable strategy for localization-based super-resolution imaging across biologically relevant volumes, bridging the gap between nanoscale precision and large-scale volumetric microscopy.
\end{abstract}

\section{Teaser}

Nanoscale imaging over large volumes in light sheet fluorescence microscopy using a twin Airy engineered point spread function.

\section{Introduction}

Nanoscale localization microscopy has transformed the study of biological systems by enabling the visualization and quantification of molecular organization beyond the diffraction limit \cite{betzig2006imaging, huang2010breaking}. Applications range from resolving the architecture of the actin cytoskeleton \cite{xu2012dual} to quantifying receptor dynamics \cite{manley2008high}  and characterizing oncogenic signaling complexes \cite{cox2012bayesian, sahl2017fluorescence}. Despite these advances, most localization measurements remain confined to two-dimensional imaging planes or limited axial ranges. Extending nanoscale localization to large three-dimensional volumes remains a fundamental challenge, particularly when high localization precision, imaging speed, and large volumetric fields of view are required simultaneously.

Single-emitter localization approaches have been at the heart of many of these advances, enabling nanometer-scale precision through diverse techniques such as engineering the emission point spread function (PSF) \cite{pavani2009three}. By encoding axial position within the PSF, three-dimensional localization becomes possible without mechanical sectioning. However, increasing the axial range of engineered PSFs (ePSFs) introduces a fundamental trade-off: as the encoded depth range expands, localization precision degrades and emitter density must be reduced to avoid PSF overlap. Consequently, methods capable of operating over axial ranges of tens of micrometers \cite{shechtman2015precise} often require sparse labeling densities or acceptance of lower localization precision.

Recent strategies have sought to overcome these constraints by computationally resolving overlapping emitters using advanced deconvolution techniques \cite{harvey2024volumetric} or deep neural networks \cite{nehme2020deepstorm3d}. As an alternative, Daly et al. \cite{daly2024high} demonstrated substantial improvements by disentangling overlapping emitters via parallax reconstruction. Nevertheless, these approaches remain limited to axial ranges of approximately 8~$\mu$m or less, underscoring a persistent challenge: nanometer-scale localization precision and extended volumetric imaging range have yet to be achieved simultaneously. Overcoming this limitation would enable nanoscale mapping throughout extended biological volumes, providing structural information across cellular and multicellular systems rather than within isolated optical sections. 

\begin{figure}[htb!]
    \centering
    \includegraphics[width=\textwidth]{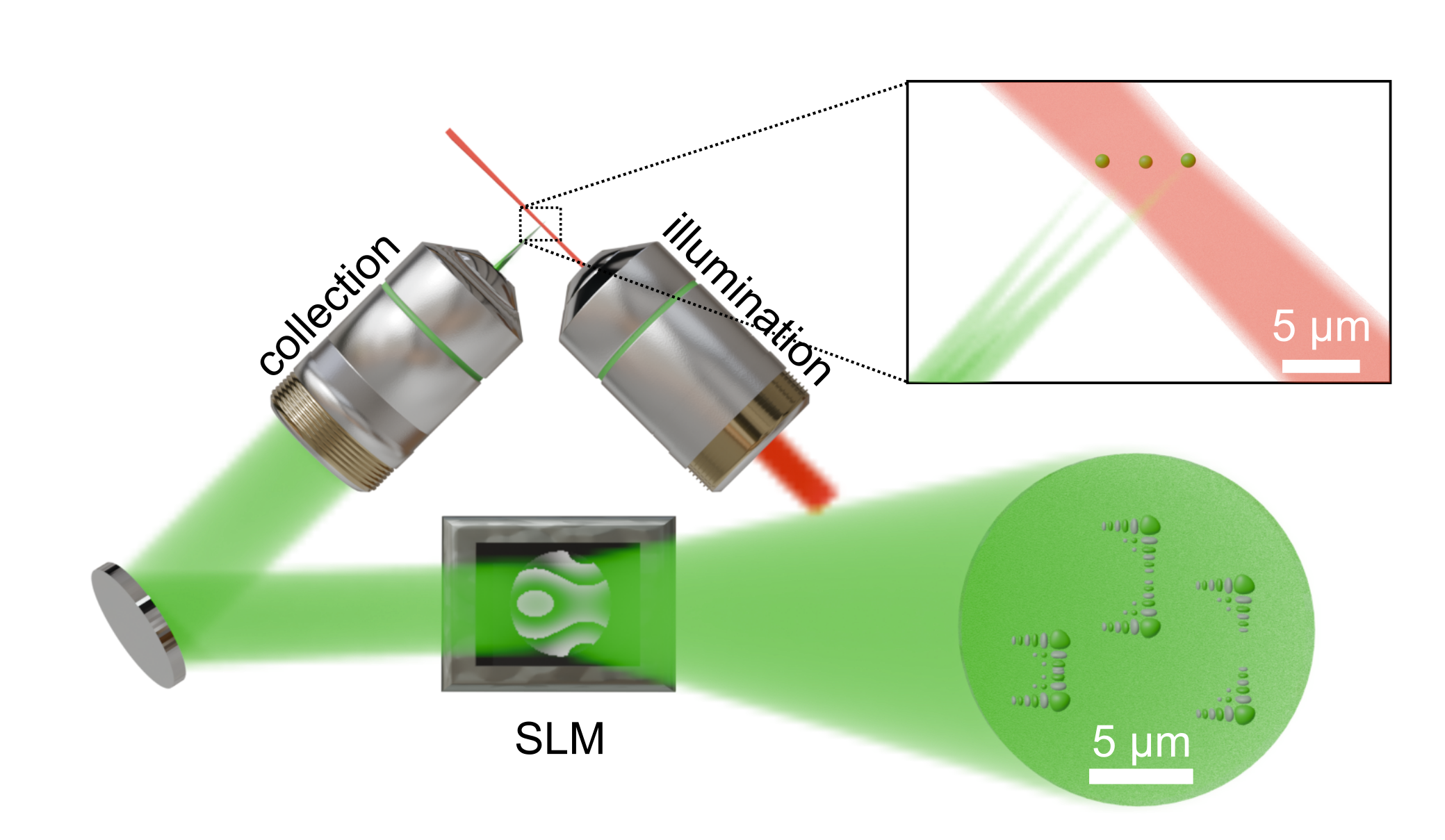}
    \caption{\textbf{ePSF-enhanced LSFM setup and localization principle.} A Gaussian light sheet selectively illuminates a cross-section of point emitters in the sample volume via an illumination objective. The fluorescence from the emitters is collected using an orthogonal collection objective. A spatial light modulator (SLM) is positioned in the Fourier plane of the collection arm, yielding a twin Airy ePSF in the camera plane, as shown in the green circle. The shape of the PSF allows the position of the emitters within the illuminated cross-section to be inferred with sub-diffraction localization precision of less than 20~nm in the lateral dimensions and 42~nm in the axial dimension.}
    \label{fig:ConceptualOverview}
\end{figure}

Here, we address the challenge of achieving nanoscale three-dimensional localization microscopy across large volumes. To achieve this goal, we judiciously combine a twin Airy ePSF \cite{zhou2020twin} with light sheet fluorescence microscopy (LSFM) \cite{gustavsson20183d}, as illustrated in Fig. \ref{fig:ConceptualOverview}. LSFM is a cornerstone of biomedical research \cite{olarte2018light}, offering rapid volumetric imaging with low photodamage \cite{verveer2007high, chow2024quantifying}. Although the axial resolution of LSFM is typically limited to a  few micrometres by its optical sectioning capability, we demonstrate nanometre-scale localization precision throughout the illuminated volume by using a twin Airy ePSF optimized for the light sheet illumination profile. By translating the sample through the light sheet, we localize emitters in volumes spanning hundreds of micrometers. Our approach surpasses both the localization precision previously achieved using LSFM and the imaging volumes previously demonstrated using ePSF nanoscale three-dimensional localization. Consequently, we achieve sub-20~nm lateral localization precision over a field of view of 295~$\mu$m $\times$ 330~$\mu$m and 42~nm axial localization precision over an axial range of 100~$\mu$m. Current performance is limited primarily by mechanical vibrations in the experimental apparatus; in the absence of these vibrations, sub-10-nm localization precision is achievable over cubic-millimeter imaging volumes.

\section{Results}

\subsection{Experimental Setup}

Our open-top two-photon light sheet geometry employs a femtosecond-pulsed Gaussian illumination beam operating at near-infrared wavelengths (860~nm for fluorescent bead localization and 790~nm for imaging studies). The beam is digitally scanned to form a light sheet with a full-width half-maximum thickness of 4.7~$\mu$m and a Rayleigh range of 150~$\mu$m. Localization of beads in this light sheet enables a field of view of 295~$\mu$m x 330~$\mu$m as it is possible to localize beads beyond the Rayleigh range of the system. The collection arm of the LSFM incorporates a spatial light modulator (SLM) positioned at the back focal plane. Fluorescence emitted by the sample is collected by the collection objective and modulated by a phase mask displayed on the SLM before being recorded by a camera. In doing so, the positional information of the emitter is encoded into the recorded intensity pattern. It is emphasized that this straightforward modification can be incorporated into a wide range of existing LSFM systems, enabling their use for single-emitter localization. Further details regarding the experimental setup can be found in the Methods.

\subsection{An optimized ePSF for single-emitter localization in a light sheet geometry}

A numerical model accurately describing the propagation of light from the sample plane to the camera plane of the experimental system is required for optimization of the microscope ePSF. The development of both a narrowband and broadband propagation model for the two-photon light sheet geometry are described in the Methods. The narrowband model is suitable for simulating the propagation of laser light, whereas the broadband model is appropriate for simulating fluorescent emission over bandwidths spanning hundreds of nanometers. Additionally, the Cramér-Rao lower bound (CRLB) \cite{ober2004localization} analysis and optimization algorithms used to generate the optimal ePSF are detailed in the Supplementary Materials.

The use of an SLM warrants particular consideration. When using an SLM, a blazed grating is typically used to separate the zero order from the first order, such that a spatial filter can then be used to eliminate the zero order completely. However, for broadband fluorescence emission (spanning approximately 150~nm in the case of our fluorescent bead samples), this induces PSF blurring (Fig. \ref{fig:BroadbandPSF}) \cite{spangenberg2014white, martens2022enabling, munagavalasa2017spatial} as different spectral components experience different phase modulations.

Typically, this effect is mitigated through spectral filtering such that only a small range of wavelengths is preserved \cite{amin2021localization, amin2022multicolor, brenner2025implementation, jiao2023simultaneous}. Using our propagation model, however, we demonstrate that preservation of the zero order does not adversely affect the localization precision (see Supplementary Material). Furthermore, spatial filtering reduces the available space-bandwidth product, while spectral filtering reduces the detected photon budget, both of which degrade performance. Consequently, we choose to preserve the zero order throughout this study.
Preserving the zero order yields a simplified implementation and preserves the space-bandwidth product and maintains a high photon flux, consistent with previous SLM-based widefield imaging approaches \cite{Upadhya2022}.

A wide variety of ePSFs have been previously considered for epifluorescence microscope geometries, including astigmatic \cite{huang2008three}, double helix \cite{pavani2009three}, tetrapod \cite{shechtman2015precise} and twin Airy ePSFs \cite{zhou2020twin}. The tetrapod and twin Airy ePSFs both provide extended axial localization ranges\cite{zhou2020twin, shechtman2015precise}, and exhibit comparable performance in our LSFM geometry (see Supplementary Materials). We adopt the twin Airy phase mask for the remainder of this study as the $z$ position of the emitter can be inferred directly from the displacement of the two principal Airy beams (Fig. 1b), providing a simple and intuitive route to three-dimensional localization \cite{zhou2020twin}.

CRLB analysis in conjunction with the propagation model was used to optimize the twin Airy ePSF for the generated light sheet (4.7~$\mu$m thickness), yielding an optimal phase-mask parameter of $\alpha=5.35$, where $\alpha$ is a scale factor that determines the peak phase modulation of the twin Airy ePSF \cite{zhou2020twin}. The resultant design achieves a theoretical localization precision of $\Delta x$ = 2.72~nm, $\Delta y$ = 3.28~nm and $\Delta z$ = 9.54~nm along the $x$, $y$ and $z$ axes, respectively.

\subsection{Imaging}

\begin{figure}
    \centering
    \includegraphics[width=\textwidth]{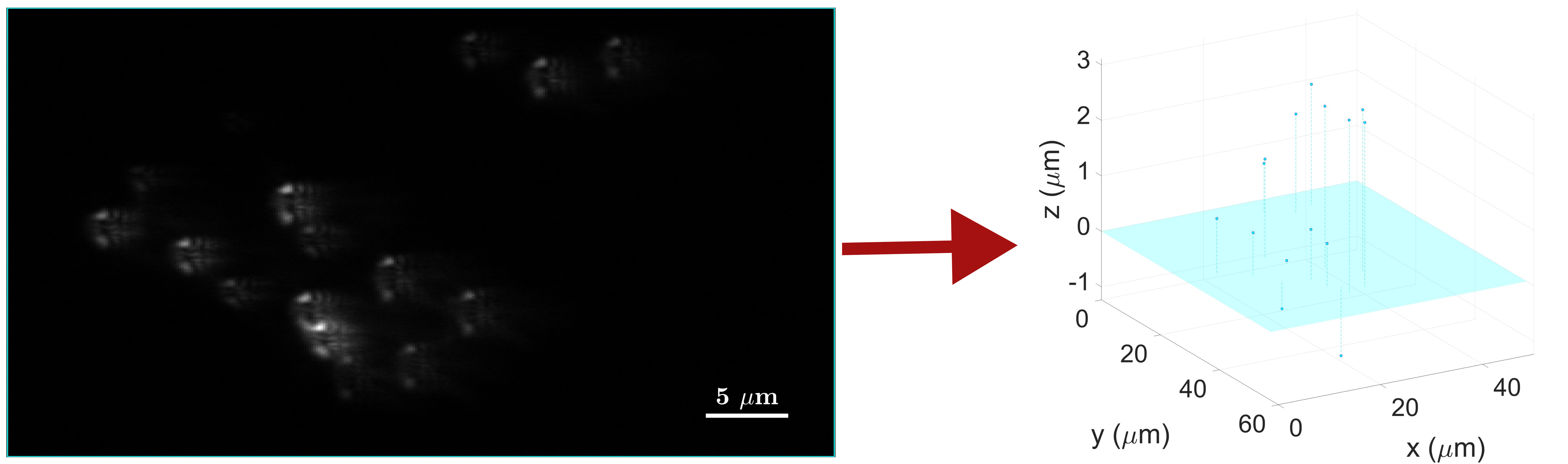}
    \caption{\textbf{Single-plane ePSF-LSFM imaging and 3D localization using twin Airy ePSF.} Example single-plane ePSF-LSFM image acquired using a 4.7~$\mu$m-thick light sheet. The sample consists of 200~nm green fluorescent beads embedded in agar, and the original field of view has been cropped for illustrative purposes. (Left) Raw acquired image showing the twin Airy ePSFs generated by individual emitters. (Right) Reconstructed bead positions obtained with localization precisions of $\Delta x = 19$~nm, $\Delta y = 16$~nm, and $\Delta z = 42$~nm. The axial bead positions are shown relative to the center of the Gaussian light sheet, indicated by the blue plane.}
    \label{fig:AgarSinglePlane}
\end{figure}

The performance of the ePSF-enabled two-photon LSFM system was evaluated using single-plane imaging of a phantom sample comprising 200 nm diameter fluorescent beads embedded in agar (Fig. \ref{fig:AgarSinglePlane}). The characteristic twin Airy ePSFs are readily resolved in the raw data, enabling direct inference of the three-dimensional emitter position relative to the focal plane.

Localization precision is quantified from repeated image acquisitions on stationary fluorescent beads dispersed across the full field of view. Estimates of $\Delta x = 19$ nm, $\Delta y = 16$ nm, and $\Delta z = 42$ nm are obtained from 1400 fluorescent bead detections. Under these conditions, the experimentally achieved precision is primarily limited by thermal drift and mechanical vibration, rather than by the intrinsic performance of the ePSF \cite{ober2004localization, backer2014extending, hess2006ultra}.

\begin{figure}
    \centering
    \includegraphics[width=\textwidth]{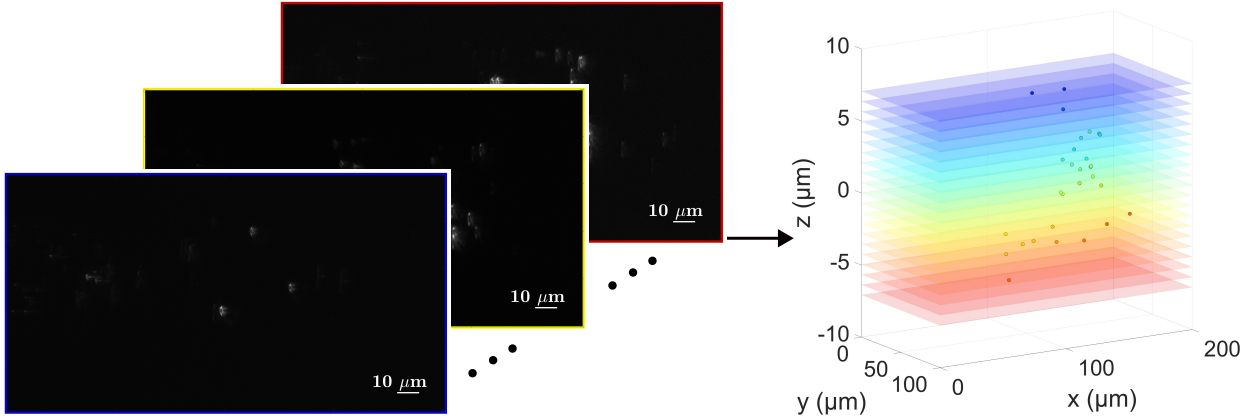}
    \caption{\textbf{Multi-plane ePSF-LSFM imaging and 3D localization using twin Airy ePSF.} Example multi-plane ePSF-LSFM imaging acquired by scanning the sample through a 4.7~$\mu$m-thick light sheet in 0.7~$\mu$m steps. The sample consists of 200~nm green fluorescent beads embedded in agar, and the original field of view has been cropped for illustrative purposes. (Left) Stack of raw acquired images showing the twin Airy ePSFs at different axial positions. (Right) Reconstructed bead positions. The colored outlines in (Left) indicate the corresponding in-focus planes shown in (Right).}
    \label{fig:AgarVolume}
\end{figure}

Scanning the sample through the light sheet allows a volumetric map of emitters to be constructed. For example, Fig. \ref{fig:AgarVolume} shows volumetric imaging over an effective axial range of 10~$\mu$m. The sample was stepped through the light sheet in increments of 0.7~$\mu$m, significantly less than its 4.7~$\mu$m thickness, and images were acquired with an exposure time of 200 ms per plane. Where necessary, larger axial ranges can be accessed by extending the scan range. 

Each emitter is detected in multiple overlapping illumination planes, enabling the relative displacement between successive planes to be determined with an uncertainty governed by the localization precision of the system. Although depth-dependent aberrations in the collection arm become increasingly significant when imaging larger volumes, their impact is mitigated by localizing emitters across multiple illumination planes.

\begin{figure}
    \centering
    \includegraphics[width=1\linewidth]{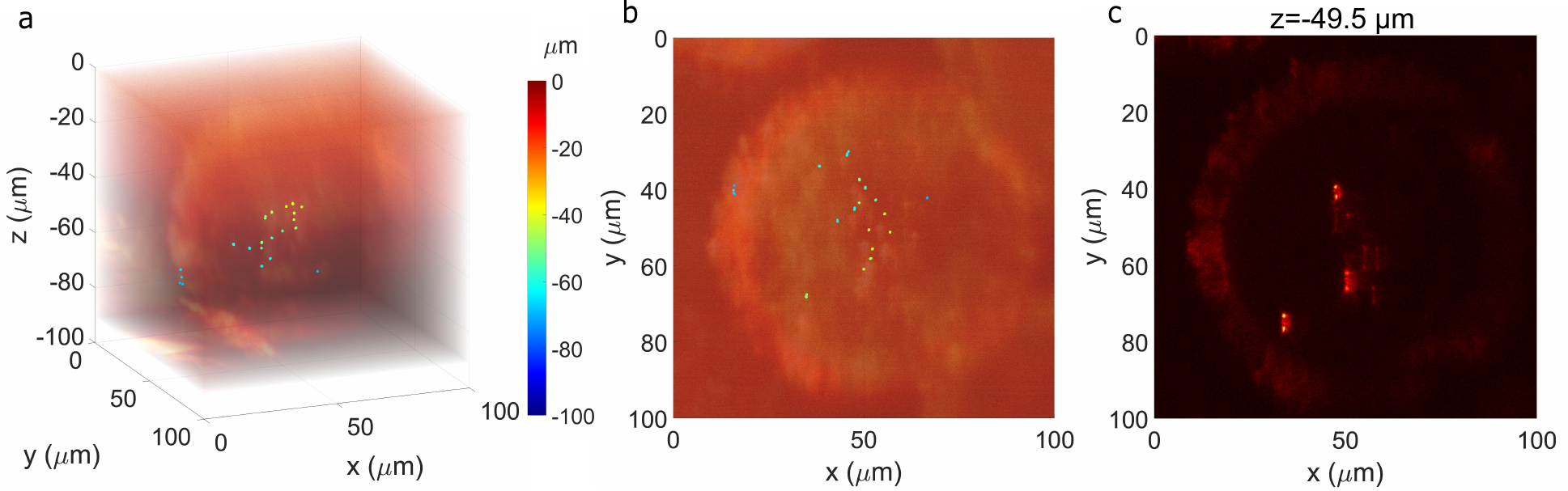}
    \caption{\textbf{Three-dimensional imaging of a murine oocyte with ePSF-based fluorescent bead localization.} a) Three-dimensional reconstruction of a murine oocyte stained with Phalloidin-AF 488 to delineate cellular boundaries, acquired using a conventional PSF, and overlaid with fluorescent bead localizations obtained using the twin Airy ePSF. Localized beads are distributed throughout the oocyte volume. b) Top-view projection of the reconstructed dataset. c) Representative optical section showing twin Airy ePSFs from localized fluorescent beads within the oocyte.}
    \label{fig:bio}
\end{figure}

To demonstrate performance in a biologically relevant setting, murine oocytes were injected with fluorescent beads and subsequently imaged, providing a more stringent test of single emitter localization within an extended volume (Fig. \ref{fig:bio}). Mammalian oocytes are particularly large cells (approximately 80~$\mu$m in diameter), presenting a significant challenge for high-resolution volumetric imaging \cite{marchais2025imaging}. 

The beads within the oocytes were imaged and localized using the optimized ePSF mask displayed on the SLM. Independently, Alexa Fluor-stained transzonal projections were imaged without an ePSF mask and using a different excitation wavelength, allowing a conventional LSFM image to be acquired. The localized bead positions were subsequently overlaid onto the conventional LSFM image, providing structural context and illustrating the flexibility of the SLM-enabled ePSF-LSFM platform.

The beads are localized with sub-20~nm lateral precision across a field of view of 295~$\mu$m $\times$ 330~$\mu$m and 42~nm axial precision over an axial range of 100~$\mu$m, although the field of view shown in Fig. \eqref{fig:bio} has been cropped for clarity. These results demonstrate nanoscale localization throughout a biologically relevant imaging volume spanning hundreds of micrometers laterally and 100~$\mu$m axially.

\section{Discussion}

We demonstrate that integrating a twin Airy ePSF with two-photon LSFM enables nanoscale localization across extended imaging volumes. By combining engineered detection with confined excitation, localization precision is maintained while scaling to substantially larger volumes than previously achieved with ePSF-based localization microscopy. Although other ePSFs (e.g., tetrapod) could be used, the twin Airy ePSF offers a simple and computationally efficient route to 3D localization \cite{zhou2020twin}.

Using established ePSF methods, extending axial range typically reduces localization precision and tightens emitter density constraints due to PSF overlap. Here, light sheet excitation confines fluorescence generation to a thin axial region \cite{gustavsson20183d}, limiting emitters per frame and suppressing overlap at acquisition. This enables high-precision localization without computational multi-emitter recovery. Consequently, instantaneous emitter density is governed by the illuminated light sheet volume rather than the full reconstructed volume.

Optical sectioning by the light sheet ensures that only $\sim $100–200 emitters occupy an illuminated volume of $\sim 4.6 \times 10^{-4}$ mm$^3$, corresponding to densities of $10^5$–$10^6$ emitters per mm$^3$. Because only emitters within the light sheet are visible within an acquired frame, larger volumes can be reconstructed via axial scanning without increased overlap. Thus, instantaneous emitter density is effectively decoupled from total imaging volume, enabling scalable volumetric localization microscopy.

The system achieves lateral localization precisions of 16–19~nm over fields of view spanning hundreds of micrometers, and axial precision of 42~nm over a 100~$\mu$m depth range. This corresponds to an increase in accessible imaging volume of several orders of magnitude compared to conventional ePSF-based approaches while maintaining comparable localization precision. In contrast to standard LSFM, which is typically limited to micrometre-scale axial resolution, the incorporation of an engineered PSF enables nanometre-scale localization within the excitation volume, effectively bridging the gap between volumetric imaging and super-resolution microscopy.

Previous volumetric precision localization methods \cite{gustavsson20183d, cheng2024light} note that the use of light sheet illumination reduces both photobleaching and background signal. While these works also combine light sheet illumination with ePSF engineering, the axial range obtained is dictated by the range of the ePSF and is limited to 10~$\mu$m. Instead, our approach breaks this constraint and the achievable imaging volume is not limited by the ePSF but by scan range, sample translation, aberrations, and system geometry. Current limits arise from mechanical vibration and stability rather than theoretical localization bounds, indicating a path toward millimeter-scale volumes and sub-10~nm precision.

This approach complements high-density volumetric super-resolution methods that computationally disentangle overlapping emitters \cite{daly2024high}. While such methods tolerate higher instantaneous densities, they require greater algorithmic complexity and computational resources. In contrast, our strategy enhances emitter separability optically at acquisition, reducing computational demands while preserving localization fidelity. 

Future improvements may leverage spatial light modulator programmability for dynamic phase-mask control, enabling correction of depth-dependent aberrations and signal variation. Enhanced light-sheet strategies could further improve excitation uniformity and reduce shadowing. More fundamentally, the present framework demonstrates how optical confinement and engineered detection can be combined to overcome conventional scaling limitations in volumetric localization microscopy. By shifting the burden of emitter separability from computational post-processing to optical control at the point of acquisition, this approach establishes a scalable route toward volumetric nanoscale imaging without requiring complex multi-emitter fitting or deep learning-based reconstruction.

Overall, this approach enables nanoscale localization across biologically relevant volumes. By combining optical confinement with engineered detection, it provides a scalable alternative to high-density computational methods, bridging nanoscale precision and large-volume imaging for 3D super-resolution microscopy.

\section{Materials and methods}

\subsection{Experimental setup}

\begin{figure}[htb]
    \centering
    \includegraphics[width=1\linewidth]{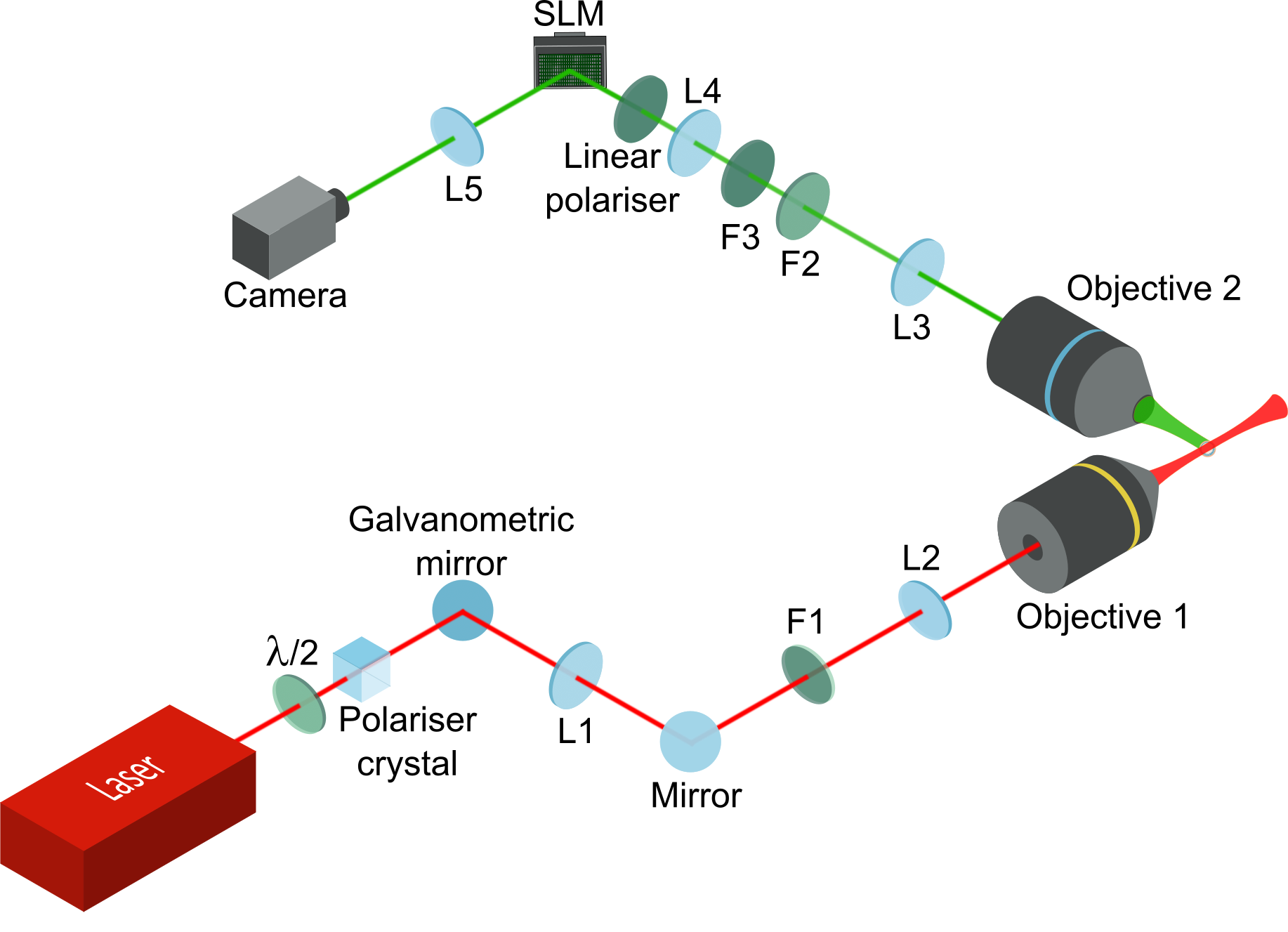}
    \caption{\textbf{Detailed schematic of the ePSF-LSFM setup.} Components used are detailed in Table \ref{tab:Setupcomponents}. The letter F denotes filters, while the letter L denotes lenses.}
    \label{fig:SetupSchematic}
\end{figure}

A conceptual overview of the experimental setup is provided in Fig. \ref{fig:ConceptualOverview}. A detailed schematic of the ePSF-LSFM setup is provided in Fig. \ref{fig:SetupSchematic}, and additional parameters describing the setup are provided in Table \ref{tab:Setupcomponents}.

An ultra-short pulse femtosecond laser (Coherent Chameleon Vision S, pulse duration of 57~fs) generates a Gaussian beam. This is scanned to form a light sheet using a galvanometric mirror and focused onto the sample using a 10x Nikon CFI Plan Fluorite objective (0.30 NA, 3.5 mm WD). The light sheet in the sample plane has a Gaussian profile with a full-width half-maximum of 4.7~$\mu$m and a Rayleigh range of 150~$\mu$m. Fluorescence emitted by a two-photon absorption process is collected with a 16x Nikon CFI LWD Plan Fluorite objective (0.80 NA, 3.0 mm WD) and modulated using an E19X12-500-1200 Meadowlark spatial light modulator (SLM) before the final image is recorded using an ORCA-Quest 2 qCMOS C15550-22UP camera. The SLM is positioned conjugate to the back focal plane of the collection objective and is used to modulate the PSF of the collection arm.

\subsection{Sample preparation - fluorescent beads in agar}

In order to characterise the twin Airy point spread function (PSF) of the imaging system, phantoms made up of 200 nm diameter green fluorescent (Ex: 860 nm/Em: 532 nm) microspheres (Invitrogen, Thermo Fisher Scientific, MA, USA) mixed with 1 $\%$ (w/v) solution of low-melting-point agarose (Invitrogen, Thermo Fisher Scientific, MA, USA) at 1:1000 concentration were prepared. The samples were manually pipetted into a 3D-printed sample holder designed for light sheet imaging as previously described in Chow et al. \cite{chow2024quantifying}. This setup provides a stable, optically clear matrix for 3D visualization of beads with minimal sample movement during acquisition.

\subsection{Sample preparation - murine oocytes}

Female (21–23 days) CBA x C57BL/6 first filial (CBAF1) generation mice were obtained from Laboratory Animal Services (University of Adelaide, Australia) and maintained on a 12 h light: 12 h dark cycle with rodent chow and water provided ad libitum. Animal ethics were approved by the Animal Ethics Committee of the University of Adelaide (M-2019-097). Oocyte collection was conducted in accordance with the Australian Code of Practice for the Care and Use of Animals for Scientific Purposes. Briefly, female mice were administered intraperitoneally (i.p.) with 5 IU of equine chorionic gonadotropin (eCG; Folligon, Braeside, VIC, Australia). At 46 h post-eCG administration, mice were culled by cervical dislocation, and the ovaries were carefully dissected. Cumulus oocyte complexes (COCs; layers of cumulus cells surrounding an oocyte) were isolated from ovaries in prewarmed Research Wash medium (ART Lab Solutions, SA, Australia) supplemented with 4 mg/mL low fatty acid bovine serum albumin (BSA, MP Biomedicals, AlbumiNZ, Auckland, NZ) using a 29-gauge insulin syringe with needle (Terumo Australia Pty Ltd., Australia). Cumulus cells were removed via enzymatic denudation by placing COCs in hyaluronidase (500 µg/mL; Sigma-Aldrich, St. Louis, MO, USA) diluted 1:1 in Research Wash medium (ART Lab Solutions, Adelaide, Australia) for 2 minutes on a heating stage set to maintain temperature at 37.5 °C. The resultant oocytes were then used for microinjection.

\subsection{Fluorescent bead microinjection}
To demonstrate the capability of the twin Airy ePSF for deep imaging, microinjection of 200-nm fluorescent beads was performed in murine oocytes. Microinjections were performed using a Nikon Eclipse TE2000-E inverse microscope (Nikon Instruments Inc.) equipped with a Tokai Hit ThermoPlate set to maintain temperature at 37.5 °C and precision micromanipulators.

The microinjection dish was prepared by placing a 10 $\mu$L drop of Research Wash medium for oocyte manipulation and another 20 $\mu$L drop containing 200 nm fluorescent beads diluted in Research Wash medium (1:500) at the centre of a 60 mm petri dish lid (Falcon, Corning, In Vitro Technologies, VIC, Australia), overlaid with paraffin oil (Merck, Germany). The dish was pre-equilibrated on a heated stage at 37 °C for at least 4 h before use. The Research Wash medium provides a physiologically pH-buffered medium for handling live oocytes.

Following enzymatic denudation, mouse oocytes were transferred into the 10 $\mu$L oocyte manipulation drop and microinjection was performed by using a holding pipette (inner diameter: 17 $\mu$m; outer diameter: 80 $\mu$m; bevel: 30°; Cook Medical, PA, USA) and injection pipette (inner diameter: 5 $\mu$m; outer diameter: 7 $\mu$m; bevel: 20°; Cook Medical) loaded onto the precision micromanipulators on the microscope. Following microinjection, live oocytes were incubated in Phalloidin-AF 488 conjugate (Invitrogen, Thermo Fisher Scientific, MA, USA) diluted to 1X in Research Wash medium for 30 minutes at at 37.5 °C in the dark. Mouse oocytes were then transferred to the 3D-printed sample holder containing a 2 $\mu$l drop of Research Wash medium overlaid with paraffin oil, as previously described in \cite{chow2024quantifying} for light sheet imaging. This enabled visualization of actin-rich transzonal projections (TZPs)---cytoplasmic extensions that span the zona pellucida to connect surrounding cumulus cells with the oocyte, thereby enabling clear delineation of oocyte boundaries and volume.


\clearpage 
\bibliography{bibliography} 

@article{shechtman2015precise,
  title = {Precise three-dimensional scan-free multiple-particle tracking over large axial ranges with tetrapod point spread functions},
  author = {Shechtman, Y. and Weiss, L. E. and Backer, A. S. and Sahl, S. J. and Moerner, W. E.},
  journal = {Nano Letters},
  volume = {15},
  number = {6},
  pages = {4194--4199},
  year = {2015}
}

@article{nehme2020deepstorm3d,
  title = {DeepSTORM3D: dense 3D localization microscopy and PSF design by deep learning},
  author = {Nehme, E. and Freedman, D. and Gordon, R. and Ferdman, B. and Weiss, L. E. and Alalouf, O. and Naor, T. and Orange, R. and Michaeli, T. and Shechtman, Y.},
  journal = {Nature Methods},
  volume = {17},
  number = {7},
  pages = {734--740},
  year = {2020}
}

@article{cheng2024light,
  title={Light sheet illumination in single-molecule localization microscopy for imaging of cellular architectures and molecular dynamics},
  author={Cheng, Siyang and Nakatani, Yuya and Gagliano, Gabriella and Saliba, Nahima and Gustavsson, Anna-Karin},
  journal={npj Imaging},
  volume={2},
  number={1},
  pages={49},
  year={2024},
  publisher={Nature Publishing Group UK London}
}

@article{daly2024high,
  title = {High-density volumetric super-resolution microscopy},
  author = {Daly, S. and Ferreira Fernandes, J. and Bruggeman, E. and Handa, A. and Peters, R. and Benaissa, S. and Zhang, B. and Beckwith, J. S. and Sanders, E. W. and Sims, R. R. and others},
  journal = {Nature Communications},
  volume = {15},
  number = {1},
  pages = {1940},
  year = {2024}
}

@article{harvey2024volumetric,
  title = {Volumetric reconstruction for improved 3D single-molecule localisation microscopy},
  author = {Harvey, A. and Olesker, D. and Taylor, J. and Handley, M.},
  journal = {Optica Open},
  year = {2024}
}

@article{zhou2020twin,
  title = {Twin-Airy point-spread function for extended-volume particle localization},
  author = {Zhou, Y. and Zammit, P. and Zickus, V. and Taylor, J. M. and Harvey, A. R.},
  journal = {Physical Review Letters},
  volume = {124},
  number = {19},
  pages = {198104},
  year = {2020}
}

@article{pavani2009three,
  title = {Three-dimensional, single-molecule fluorescence imaging beyond the diffraction limit by using a double-helix point spread function},
  author = {Pavani, S. R. P. and Thompson, M. A. and Biteen, J. S. and Lord, S. J. and Liu, N. and Twieg, R. J. and Piestun, R. and Moerner, W. E.},
  journal = {Proceedings of the National Academy of Sciences},
  volume = {106},
  number = {9},
  pages = {2995--2999},
  year = {2009}
}

@article{shechtman2014optimal,
  title = {Optimal point spread function design for {3D} imaging},
  author = {Shechtman, Y. and Sahl, S. J. and Backer, A. S. and Moerner, W. E.},
  journal = {Physical Review Letters},
  volume = {113},
  number = {13},
  pages = {133902},
  year = {2014}
}

@article{gustavsson20183d,
  title = {3D single-molecule super-resolution microscopy with a tilted light sheet},
  author = {Gustavsson, A.-K. and Petrov, P. N. and Lee, M. Y. and Shechtman, Y. and Moerner, W. E.},
  journal = {Nature Communications},
  volume = {9},
  number = {1},
  pages = {123},
  year = {2018}
}

@article{xu2012dual,
  title = {Dual-objective STORM reveals three-dimensional filament organization in the actin cytoskeleton},
  author = {Xu, K. and Babcock, H. P. and Zhuang, X.},
  journal = {Nature Methods},
  volume = {9},
  number = {2},
  pages = {185--188},
  year = {2012}
}

@article{manley2008high,
  title = {High-density mapping of single-molecule trajectories with photoactivated localization microscopy},
  author = {Manley, S. and Gillette, J. M. and Patterson, G. H. and Shroff, H. and Hess, H. F. and Betzig, E. and Lippincott-Schwartz, J.},
  journal = {Nature Methods},
  volume = {5},
  number = {2},
  pages = {155--157},
  year = {2008}
}

@article{cox2012bayesian,
  title = {Bayesian localization microscopy reveals nanoscale podosome dynamics},
  author = {Cox, S. and Rosten, E. and Monypenny, J. and Jovanovic-Talisman, T. and Burnette, D. T. and Lippincott-Schwartz, J. and Jones, G. E. and Heintzmann, R.},
  journal = {Nature Methods},
  volume = {9},
  number = {2},
  pages = {195--200},
  year = {2012}
}

@article{sahl2017fluorescence,
  title = {Fluorescence nanoscopy in cell biology},
  author = {Sahl, S. J. and Hell, S. W. and Jakobs, S.},
  journal = {Nature Reviews Molecular Cell Biology},
  volume = {18},
  number = {11},
  pages = {685--701},
  year = {2017}
}

@article{huang2010breaking,
  title = {Breaking the diffraction barrier: super-resolution imaging of cells},
  author = {Huang, B. and Babcock, H. and Zhuang, X.},
  journal = {Cell},
  volume = {143},
  number = {7},
  pages = {1047--1058},
  year = {2010}
}

@article{betzig2006imaging,
  title = {Imaging intracellular fluorescent proteins at nanometer resolution},
  author = {Betzig, E. and Patterson, G. H. and Sougrat, R. and Lindwasser, O. W. and Olenych, S. and Bonifacino, J. S. and Davidson, M. W. and Lippincott-Schwartz, J. and Hess, H. F.},
  journal = {Science},
  volume = {313},
  number = {5793},
  pages = {1642--1645},
  year = {2006}
}

@article{chow2024quantifying,
  title = {Quantifying DNA damage following light sheet and confocal imaging of the mammalian embryo},
  author = {Chow, D. J. X. and Schartner, E. P. and Corsetti, S. and Upadhya, A. and Morizet, J. and Gunn-Moore, F. J. and Dunning, K. R. and Dholakia, K.},
  journal = {Scientific Reports},
  volume = {14},
  number = {1},
  pages = {20760},
  year = {2024}
}

@article{verveer2007high,
  title = {High-resolution three-dimensional imaging of large specimens with light sheet-based microscopy},
  author = {Verveer, P. J. and Swoger, J. and Pampaloni, F. and Greger, K. and Marcello, M. and Stelzer, E. H. K.},
  journal = {Nature Methods},
  volume = {4},
  number = {4},
  pages = {311--313},
  year = {2007}
}

@article{olarte2018light,
  title = {Light-sheet microscopy: a tutorial},
  author = {Olarte, O. E. and Andilla, J. and Gualda, E. J. and Loza-Alvarez, P.},
  journal = {Advances in Optics and Photonics},
  volume = {10},
  number = {1},
  pages = {111--179},
  year = {2018}
}

@article{spangenberg2014white,
  title = {White light wavefront control with a spatial light modulator},
  author = {Spangenberg, D.-M. and Dudley, A. and Neethling, P. H. and Rohwer, E. G. and Forbes, A.},
  journal = {Optics Express},
  volume = {22},
  number = {11},
  pages = {13870--13879},
  year = {2014}
}

@article{martens2022enabling,
  title = {Enabling spectrally resolved single-molecule localization microscopy at high emitter densities},
  author = {Martens, K. J. A. and Gobes, M. and Archontakis, E. and Brillas, R. R. and Zijlstra, N. and Albertazzi, L. and Hohlbein, J.},
  journal = {Nano Letters},
  volume = {22},
  number = {21},
  pages = {8618--8625},
  year = {2022}
}

@article{amin2021localization,
  title = {Localization precision in chromatic multifocal imaging},
  author = {Amin, M. J. and Petry, S. and Shaevitz, J. W. and Yang, H.},
  journal = {Journal of the Optical Society of America B},
  volume = {38},
  number = {10},
  pages = {2792--2798},
  year = {2021}
}

@article{munagavalasa2017spatial,
  title = {Spatial and spectral imaging of point-spread functions using a spatial light modulator},
  author = {Munagavalasa, S. and Schroeder, B. and Hua, X. and Jia, S.},
  journal = {Optics Communications},
  volume = {404},
  pages = {51--54},
  year = {2017}
}

@article{brenner2025implementation,
  title = {Implementation and calibration of spectroscopic single-molecule localization microscopy},
  author = {Brenner, B. and Yeo, W.-H. and Lee, Y. and Kweon, J. and Sun, C. and Zhang, H. F.},
  journal = {BMC Methods},
  volume = {2},
  number = {1},
  pages = {2},
  year = {2025}
}

@article{amin2022multicolor,
  title = {Multicolor multifocal 3D microscopy using in situ optimization of a spatial light modulator},
  author = {Amin, M. J. and Zhao, T. and Yang, H. and Shaevitz, J. W.},
  journal = {Scientific Reports},
  volume = {12},
  number = {1},
  pages = {16343},
  year = {2022}
}

@article{jiao2023simultaneous,
  title = {Simultaneous multi-plane imaging light-sheet fluorescence microscopy for simultaneously acquiring neuronal activity at varying depths},
  author = {Jiao, Z. and Zhou, Z. and Chen, Z. and Xie, J. and Mu, Y. and Du, J. and Fu, L.},
  journal = {Optica},
  volume = {10},
  number = {2},
  pages = {239--247},
  year = {2023}
}

@article{huang2008three,
  title = {Three-dimensional super-resolution imaging by stochastic optical reconstruction microscopy},
  author = {Huang, B. and Wang, W. and Bates, M. and Zhuang, X.},
  journal = {Science},
  volume = {319},
  number = {5864},
  pages = {810--813},
  year = {2008}
}

@article{ober2004localization,
  title = {Localization accuracy in single-molecule microscopy},
  author = {Ober, R. J. and Ram, S. and Ward, E. S.},
  journal = {Biophysical Journal},
  volume = {86},
  number = {2},
  pages = {1185--1200},
  year = {2004}
}

@article{backer2014extending,
  title = {Extending single-molecule microscopy using optical Fourier processing},
  author = {Backer, A. S. and Moerner, W. E.},
  journal = {The Journal of Physical Chemistry B},
  volume = {118},
  number = {28},
  pages = {8313--8329},
  year = {2014}
}

@article{hess2006ultra,
  title = {Ultra-high resolution imaging by fluorescence photoactivation localization microscopy},
  author = {Hess, S. T. and Girirajan, T. P. K. and Mason, M. D.},
  journal = {Biophysical Journal},
  volume = {91},
  number = {11},
  pages = {4258--4272},
  year = {2006}
}

@article{marchais2025imaging,
  title = {Imaging transzonal projections in the cumulus--oocyte complexes: challenges and solutions},
  author = {Marchais, M. and Bastien, A. and Nenonene, E. K. and Khandjian, E. W. and Gilbert, I. and Robert, C.},
  journal = {Biology of Reproduction},
  volume = {113},
  number = {6},
  pages = {1414--1432},
  year = {2025}
}

@article{crocker1996methods,
  title={Methods of digital video microscopy for colloidal studies},
  author={Crocker, John C. and Grier, David G.},
  journal={Journal of Colloid and Interface Science},
  volume={179},
  number={1},
  pages={298--310},
  year={1996},
  publisher={Elsevier}
}

@inproceedings{Upadhya2022,
  author    = {Avinash Upadhya and Tienan Xu and Junxiang Zhang and Yean Jin Lim and Anna Linnenberger and Naomi Mitchell and Leonie Quinn and Elizabeth E. Gardiner and Hari Shroff and Woei Ming Lee},
  title     = {Full-field imaging contrast with spatial interpixel light modulation for light sheet},
  booktitle = {Biophotonics Congress: Biomedical Optics 2022 (Translational, Microscopy, OCT, OTS, BRAIN)},
  pages      = {MW1A.5},
  publisher  = {Optica Publishing Group},
  year       = {2022}
}

@misc{blair2008matlab,
  title={The matlab particle tracking code repository},
  author={Blair, Daniel and Dufresne, Eric},
  year={2008}
}
\bibliographystyle{sciencemag}


\section*{Acknowledgments}

All of the authors would like to acknowledge Dr. Ori Cohen and Prof. Yoav Shechtman for sharing their code for generating tetrapod PSFs.

\paragraph*{Funding:}
K.D. acknowledges support from the Australian Research Council (FL210100099).
This research was supported by the Australian Research Council Centre of Excellence in Optical Microcombs for Breakthrough Science (project number CE230100006) and funded by the Australian Government.
K.R.D. acknowledges support from the National Health and Medical Research Council (APP2029067). K.R.D. is also supported by a Future Fellowship from the Australian Research Council (FT240100291).
R.E.B.G. acknowledges support from the Australian Government Research Training Program (RTP) scholarship and from the Adelaide University and Lastek Pty Ltd through the Industry Doctoral Training Centre scholarship.

\paragraph*{Author contributions:}
Conceptualization: R.P.M. and K.D.
Experimental design and build, and experimental data acquisition: R.E.B.G.
Oocyte experiment design and execution: D.J.X.C. and K.R.D.
Coding: R.E.B.G. and R.P.M.
Writing—original draft: R.E.B.G., R.P.M. and K.D.
Writing—review and editing: R.E.B.G., R.P.M., A.U., D.J.X.C., K.R.D. and K.D.
Supervision: R.P.M., A.U., K.R.D. and K.D.
Resources: K.R.D. and K.D.
Project administration: K.D.

\paragraph*{Competing interests:}
There are no competing interests to declare.

\paragraph*{Data and materials availability:}
All data and code required to reproduce the results presented in this study have been deposited in a Figshare collection (DOI: 10.25909/c.8535393).

The Figshare collection can be provided by K. D. pending scientific review and a completed material transfer agreement. Requests for Figshare collection should be submitted to: kishan.dholakia@adelaide.edu.au

\subsection*{Supplementary materials}
Supplementary Text\\
Figs. S1 to S7\\
Tables S1 to S2\\
References \textit{(1-\arabic{enumiv})}\\ 


\newpage


\renewcommand{\thefigure}{S\arabic{figure}}
\renewcommand{\thetable}{S\arabic{table}}
\renewcommand{\theequation}{S\arabic{equation}}
\renewcommand{\thepage}{S\arabic{page}}
\setcounter{figure}{0}
\setcounter{table}{0}
\setcounter{equation}{0}
\setcounter{page}{1} 


\begin{center}
\section*{Supplementary Materials for\\ \scititle}

	R.~E.~Bautista~Gonzalez$^{1}$,
	R.~Mouthaan$^{1}$,
	A.~Upadhya$^{1,2}$,\and
    D.~J.~X.~Chow$^{1,2}$,
    K.~R. Dunning$^{1,2}$,
    K.~Dholakia$^{1,3\ast}$ \and
	\small$^{1}$Centre of Light for Life and School of Biological Sciences, Adelaide University, SA 5005, Australia.\and
	\small$^{2}$Robinson Research Institute, School of Pharmacy and Biomedical Science, Adelaide University, SA 5005, Australia.\and
    \small$^{3}$SUPA, School of Physics and Astronomy, University of St Andrews, North Haugh, Fife, KY16 9SS, UK.\and
	\small$^\ast$Corresponding author. Email: kishan.dholakia@adelaide.edu.au\and
\end{center}

\subsubsection*{This PDF file includes:}
Supplementary Text\\
Figures S1 to S7\\
Tables S1 to S4\\
References \textit{(1-\arabic{enumiv})}\\ 

\newpage


\subsection*{Supplementary Text}

\subsubsection*{Narrowband \& broadband propagation models}

\begin{figure}[htb]
    \centering
    \includegraphics[width=1\linewidth]{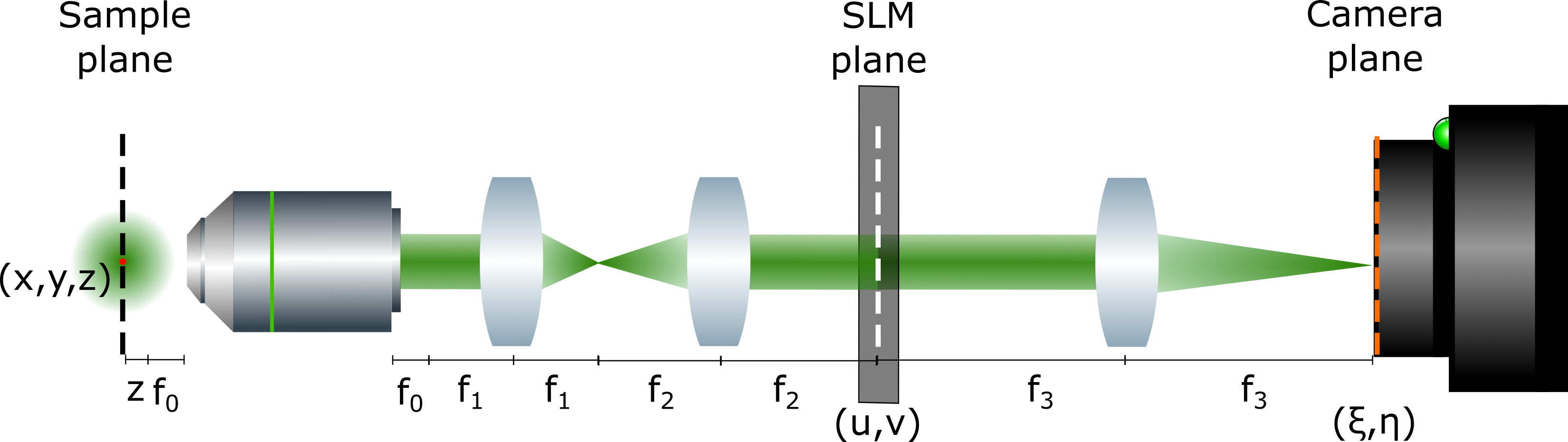}
    \caption{\textbf{Schematic for propagation model.} The fluorescence propagates from the point emitter through the 4f system and SLM to the camera.}
    \label{fig:MathsSetup}
\end{figure}

A computational model describing light propagation from the emitter to the SLM plane is required to guide the design of the engineered PSF mask. A point-source embedded in a medium with refractive index $n$ at position $(x, y, z)$ relative to the focus of the collection objective is considered. The field incident on the SLM $F_{inc}$ can then be calculated as

\begin{equation}
    F_{inc}(u, v, \lambda) = K_x K_y K_z A
\end{equation}

\noindent where $K_x$ and $K_y$ describe the effect of the transverse offset of the point-source,

\begin{align}
    K_x &= \exp(ik \cdot x \cdot u/f_\textit{eff}) & K_y &= \exp(ik \cdot y \cdot v/f_\textit{eff})
\end{align}

\noindent $K_z$ describes the effect of the axial offset of the point-source and is derived from the angular spectrum method equation,

\begin{equation}
    K_z = \exp\left( i k_n z \sqrt{1 - \left(\frac{r}{f_\textit{eff}~n}\right)^2} \right)
\end{equation}

\noindent and $A$ describes the size of the back aperture of the microscope objective, i.e. $A=1$ for $\rho < 1$ and $A = 0$ for $\rho > 1$. Variables introduced include the wavenumbers $k = \frac{2\pi}{\lambda}$, $k_n = \frac{k}{n}$ and the wavelength $\lambda$. The coordinates in the SLM plane are denoted $u$ and $v$, and the radius $r = \sqrt{u^2 + v^2}$ and normalised radius $\rho = \frac{r}{(f_{eff} \cdot NA)}$ are defined. Finally, $f_\textit{eff}$ corresponds to the effective focal length of optical system, which is given by $f_\textit{eff} = \frac{(f_3 \cdot f_2)}{(f_1 \cdot M)}$ where $M$ is the magnification indicated on the objective. The focal lengths $f_1$, $f_2$ and $f_3$ are associated with lenses indicated in Fig. \ref{fig:MathsSetup} and Table \eqref{tab:Setupcomponents}. The numerical aperture of the microscope objective in this system is not sufficiently large to demand consideration of distortions to $A$ or a vectorial model.

A phase mask displayed on the SLM imparts a spatially varying phase delay $\phi(u, v, \lambda)$ on the incident beam. The zero order field in the camera plane $F_0(\xi, \eta, \lambda)$ is calculated from the Fourier transform of the unmodulated beam, $F_0 = \mathcal{F}(F_{inc})$ whereas the first order field in the camera plane $F_1(\xi, \eta, \lambda)$ is calculated from the Fourier transform of the modulated beam $F_1 = \mathcal{F}(F_{inc} e^{i\phi})$. The field in the camera plane can then be calculated from the coherent sum of the zero and first order $\omega_0 F_0 + \omega_1 F_1$, weighted by  factors $\omega_0$ and $\omega_1$ that account for the relative efficiency with which light is coupled into the zero and first orders ($\omega_0^2 + \omega_1^2 = 1$).

The incident light originates from a fluorescence process, and so is broadband in nature, whereas the mathematical model presented so far is narrowband. To account for this, the narrowband simulation is performed at evenly spaced wavelengths spanning the emission spectrum of the fluorophore. In each case, a first-order correction factor is applied to the phase imparted by the phase mask to account for the difference in effective optical path length. Explicitly, $\phi(u, v, \lambda_i) = \phi(u, v, \lambda_0) \cdot \frac{\lambda_i}{\lambda_0}$ where $\lambda_i$ is the wavelength being considered and $\lambda_0$ is the calibration wavelength of the SLM, chosen here to coincide with the peak emission wavelength of the fluorescent beads.  The intensity of the broadband PSF $I_\text{broad}$ can be calculated using an incoherent weighted sum where the emission intensity $\Omega_\lambda$ of the fluorophore is used as a weighting term.

The light sheet intensity has a Gaussian profile in the axial direction. A weighting factor $\Omega_{2p}(z) = \exp(-2z^2/\sigma^2)$ is used to account for this, where $2\sqrt{2ln(2)}\sigma$ is the full-width half-maximum beam-waist.

The PSF is scaled to yield a total photon count that agrees with experimental observations. A scaling factor $\Omega_\#$ is introduced, which is determined by measuring the total number of photons captured from an in-focus bead when the PSF of the microscope is unaltered, i.e. when no phase mask is shown on the SLM.

To summarise, the image detected by the camera is calculated by incorporating weighting factors to account for the broadband nature of the light, the Gaussian shape of the light sheet, and the overall observed photon budget. Finally, an offset term is included to account for camera bias. In this manner, we define both a narrowband model for the PSF detected by the camera,

\begin{equation}
    I_\text{narrow}(\xi, \eta) = \Omega_\# \Omega_\text{2p}(z) |\omega_0 F_0(\lambda) + \omega_1 F_1(\lambda)|^2 + \text{Offset}
\end{equation}

\noindent as well as a broadband model,

\begin{equation}
    I_\text{broad}(\xi, \eta) = \Omega_\# \Omega_\text{2p}(z) \sum_\lambda \Omega_\lambda(\lambda) |\omega_0 F_0(\lambda) + \omega_1 F_1(\lambda)|^2 + \text{Offset}
\end{equation}

\begin{figure}[htb]
    \centering
    \includegraphics[width=1\linewidth]{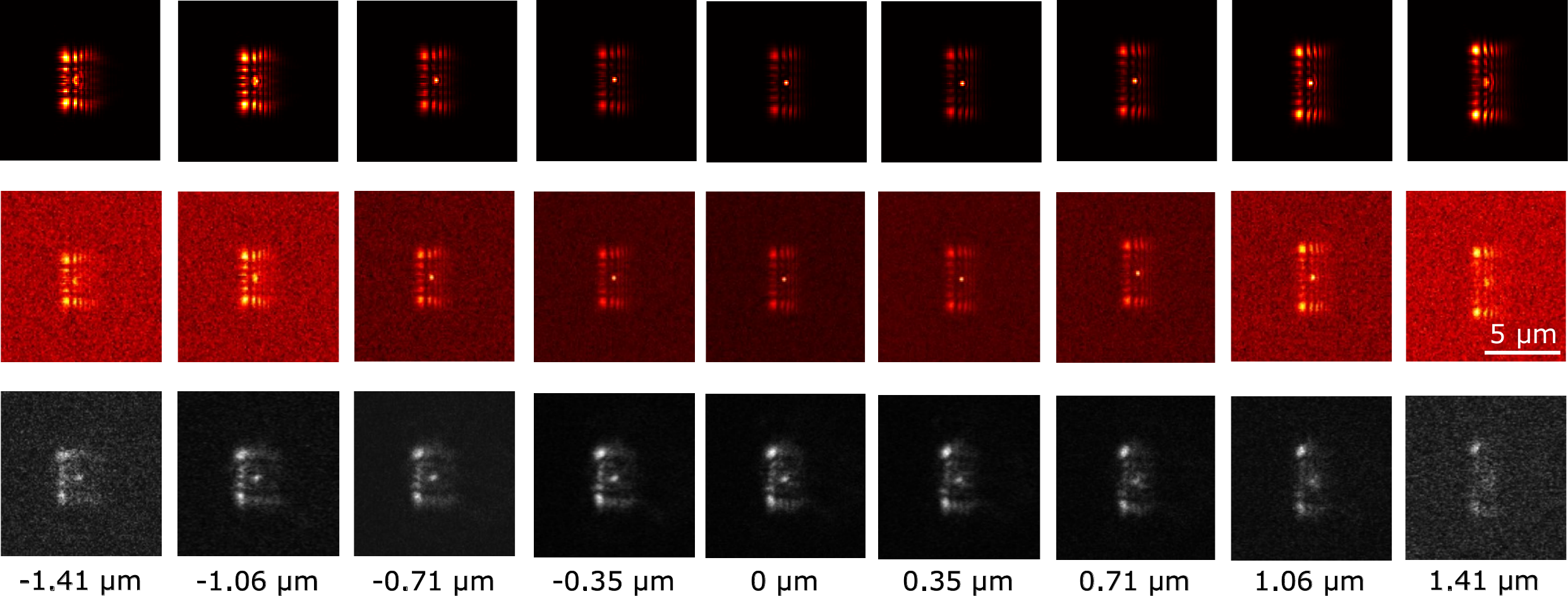}
    \caption{\textbf{Comparison of experimental vs simulated results.} The twin Airy ePSF has been simulated for various distances to the focal plane with the broadband model without noise (top row), with the broadband model with noise (middle row), and has also been experimentally characterised (bottom row).}
    \label{fig:ModelExperimentComparison}
\end{figure}

To evaluate the performance of these models, a twin Airy mask with $\alpha = 10$ is displayed on the SLM (see later for further details), and the ePSF of a single bead positioned at different axial positions is observed. This is then compared to both broadband simulations with and without noise of a bead positioned at these same depths, as illustrated in Fig. \ref{fig:ModelExperimentComparison}. Excellent agreement is obtained between all three, indicating that the broadband model is accurately representing the behaviour of the experimental system, and that this model can be relied on for PSF optimisation.

\subsubsection*{Fisher information, Cram\'er-Rao lower bound \& localization precision}

The theoretical limit on the localization precision is given by the square root of the Cram\'er-Rao lower bound (CRLB), where the CRLB is the limit on the variance of an unbiased estimator of a given parameter, derived as follows. 

The amount of information regarding an unknown parameter $\theta$ encoded into the PSF is quantified by the Fisher information $\text{FI}$, defined as

\begin{equation}
    \text{FI} = \mathbf{E} \left( \left[\frac{\partial}{\partial \theta} \log(P(X|\theta)) \right]^2 \right)
\end{equation}

\noindent Here, $P(X|\theta)$ is the probability function associated with observing $X$ given $\theta$. $\log(P(X|\theta))$ is the log-likelihood, and $\frac{\partial}{\partial \theta} \log(P(X|\theta))$ is the score. Assuming a Poissonian noise model, the following expression is obtained

\begin{equation}
    \text{FI}^\text{Poissonian}_{\theta_a, \theta_b} =\sum_{\xi, \eta} \frac{1}{I} \frac{\partial I}{\partial \theta_a} \frac{\partial I}{\partial \theta_b}
\end{equation}

Alternatively, assuming a Gaussian noise model, the following expression is obtained

\begin{equation}
    \text{FI}^\text{Gaussian}_{\theta_a, \theta_b} =\sum_{\xi, \eta} \frac{1}{\sigma^2} \frac{\partial I}{\partial \theta_a} \frac{\partial I}{\partial \theta_b}
\end{equation}

\noindent where $\sigma$ is the standard deviation of the noise. The derivatives in these equations can be evaluated using a finite difference approach which allows the so-called Fisher information matrix $\text{FI}_{mat}$ to be calculated

\begin{equation}
    \text{FI}_{mat} = 
    \begin{pmatrix}
    FI_{xx} & FI_{xy} & FI_{xz} \\
    FI_{yx} & FI_{yy} & FI_{yz} \\
    FI_{zx} & FI_{zy} & FI_{zz} \\
    \end{pmatrix}
\end{equation}

The CRLB matrix $\text{CRLB}_{mat}$ is calculated from the inverse of $FI_{mat}$, such that

\begin{equation}
    \text{CRLB}_{mat} = |\text{FI}_{mat}^{-1}| =     
    \begin{pmatrix}
    \text{CRLB}_{xx} & \text{CRLB}_{xy} & \text{CRLB}_{xz} \\
    \text{CRLB}_{yx} & \text{CRLB}_{yy} & \text{CRLB}_{yz} \\
    \text{CRLB}_{zx} & \text{CRLB}_{zy} & \text{CRLB}_{zz} \\
    \end{pmatrix}
\end{equation}

The localization precision along the $x$, $y$ and $z$ dimensions is then given by the diagonal elements $\sqrt{\text{CRLB}_{xx}}$, $\sqrt{\text{CRLB}_{yy}}$ and $\sqrt{\text{CRLB}_{zz}}$ respectively.

\subsubsection*{PSF optimisation}

Calculation of the CRLB for PSF engineering has been previously discussed in the literature \cite{ober2004localization}, and expressions for the CRLBs associated with the $x$, $y$ and $z$ positions were derived in the previous section.

The optimal PSF minimises the localization precision along $x$, $y$ and $z$ over a given axial range of interest. The axial range of interest is divided into regular intervals, and the sum of the localization precisions in $x$, $y$ and $z$ across these intervals is then used as an error metric $\text{Err}$ that is to be minimised. Matlab's nonlinear `fmincon' solver is used to perform the optimisation, as also previously done by Schechtman et al. \cite{shechtman2014optimal}:

\begin{equation}
    \text{Err}(z) = \sum_{z=min}^{z=max} (\text{CRLB}_{xx}(z) + \text{CRLB}_{yy}(z) + \text{CRLB}_{zz}(z)
\end{equation}

The less computationally intensive narrowband model is used when optimising the PSF, whereas the broadband model is used to evaluate the performance of the final PSF.

\begin{table}
    \centering
    \caption{\textbf{Comparison of localization precision obtained using narrowband model, and localization precision obtained using broadband model.} $\alpha = $ weighting of twin Airy ePSF; $\Delta x$ = localization precision in $x$; $\Delta y$ = localization precision in $y$; $\Delta z$ = localization precision in $z$. Simulations performed assuming the ePSF light sheet configuration and 50,000 detected photons.}
    \label{tab:PSF_BroadbandNarrowband}
    \begin{tabular}{c c | c c c}
        & & $\Delta x$ (nm) & $\Delta y$ (nm) & $\Delta z$ (nm) \\
        \hline
        $\alpha$ = 5.0 & Narrowband & 2.56 & 2.87 & 8.57 \\
        $\alpha$ = 5.0 & Broadband & 2.74 & 3.24 & 9.27 \\
        \hline
        $\alpha$ = 7.5 & Narrowband & 2.59 & 3.23 & 10.26 \\
        $\alpha$ = 7.5 & Broadband & 2.94 & 3.88 & 12.01 \\
        \hline
        $\alpha$ = 10.0 & Narrowband & 2.97 & 3.74 & 11.83 \\
        $\alpha$ = 10.0 & Broadband & 3.56 & 4.92 & 15.23 \\
    \end{tabular}
\end{table}

Table \ref{tab:PSF_BroadbandNarrowband} provides a comparison of the twin Airy ePSF localization precision evaluated using the narrowband and broadband models. It is observed that the narrowband propagation model yields a reasonable approximation of the more accurate broadband model for low $\alpha$ values. Larger $\alpha$ values yield masks with steeper phase gradients, which can lead to blurring of the PSF. This effect is not captured by the narrowband model, which then moderately underestimates the achieved localization precision. In this work, the narrowband model is used as part of the PSF optimisation procedure, and the final localization precision is evaluated using the broadband model.

\subsubsection*{Modulating Fluorescence using a Liquid Crystal SLM}

\begin{figure}
    \centering
    \includegraphics[width=0.6\textwidth]{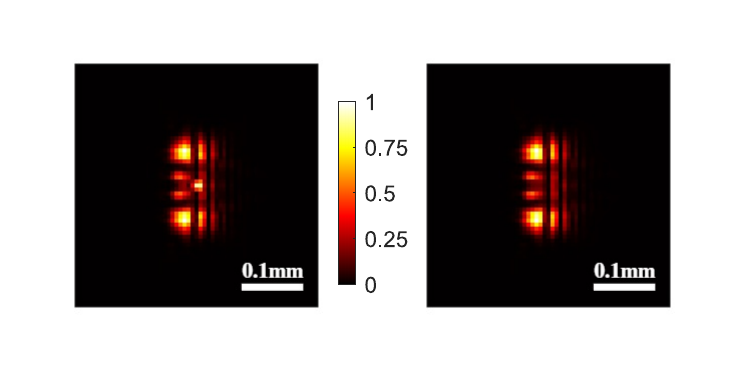}
    \caption{\textbf{Effect of zero order on the $\mathbf{\boldsymbol{\alpha} = 5.35}$ PSF.} (Left) PSF with 18\% of power in zero order, achieving a mean localization precision of 2.72~nm, 3.28~nm and 9.54~nm in the $x$, $y$ and $z$ directions respectively. (Right) PSF without zero order, achieving a mean localization precision of 2.70~nm, 3.38~nm and 9.59~nm in the $x$, $y$ and $z$ directions respectively.}
    \label{fig:PSF_ZeroOrder}
\end{figure}

Approximately 82\% of the incident light is modulated by the SLM. The remainder either reflects off the front surface of the SLM or is incident on the dead space between pixels, forming a zero order (see Fig. \ref{fig:PSF_ZeroOrder}). The zero order forms a Gaussian PSF that is superimposed with the first order twin Airy ePSF. Previous studies employ a refractive phase mask made from a material such as glass \cite{gustavsson20183d} to avoid this issue. While a glass phase mask is more photon efficient, it does not provide the same degree of flexibility as an SLM.

\begin{figure}[htb!]
    \centering
    \includegraphics[width=0.8\textwidth]{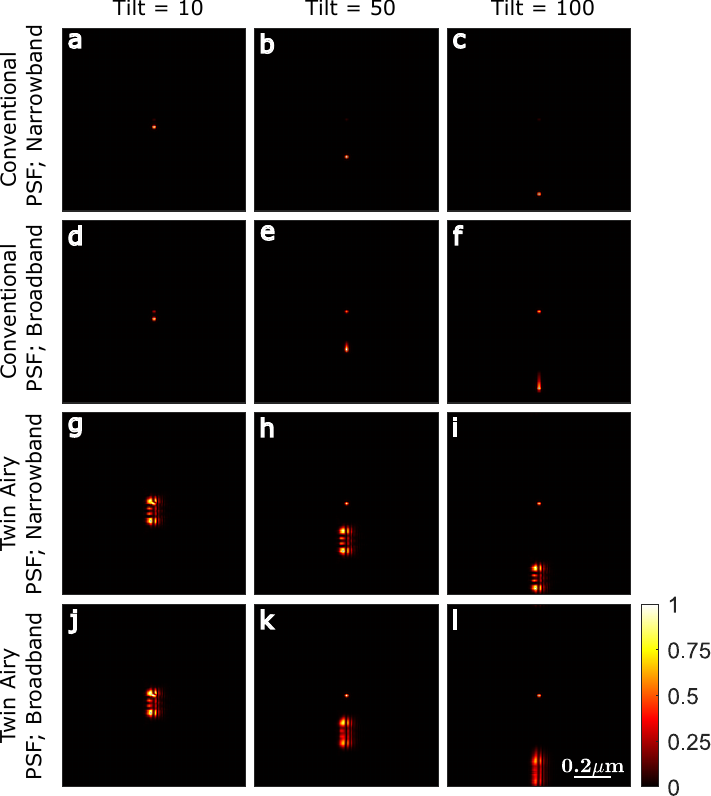}
    \caption{\textbf{Simulated effect of blazed grating on conventional Gaussian PSF and twin Airy ePSF.} Simulated effect of blazed grating on conventional Gaussian PSF assuming a narrowband signal (a, b and c) and assuming a broadband signal (d, e and f). Simulated effect of blazed grating on twin Airy ePSF assuming a narrowband signal (g, h and i) and assuming a broadband signal (j, k and l). Simulations performed assuming the ePSF light sheet configuration and an $\alpha=10$ twin Airy ePSF. Blurring effects become more evident for broadband signals when a large phase ramp is used. No blurring effects are observed when a narrowband signal is considered.}
    \label{fig:BroadbandPSF}
\end{figure}

When using an SLM, a blazed grating is typically used to separate the zero order from the first order, such that a spatial filter can then be used to eliminate the zero order completely. The steep phase gradient required to achieve this leads to a blurring of the PSF (Fig. \ref{fig:BroadbandPSF}). The root cause of this is that the fluorescence being modulated is broadband, spanning approximately 150~nm in the case of the fluorescent bead samples, and each constituent wavelength undergoes a different modulation. Previous studies have eliminated this issue by demonstrating a proof of concept using coherent light \cite{pavani2009three, backer2014extending}, making the approach less relevant for biological studies. Alternatively, a narrowband spectral filter has been used to select a single wavelength \cite{amin2021localization, amin2022multicolor, brenner2025implementation, jiao2023simultaneous}, reducing the number of photons detected and hence the localization precision.

To investigate the impact of the presence of the zero order, the localization precision associated with a twin Airy ePSF is evaluated for the case where 18\% of power is coupled into the zero order (corresponding to an SLM being used) and for the case where all power is coupled into the first order (corresponding to a phase plate being used). This is done using the broadband model, with results shown in Table \ref{tab:PSF_ZeroOrder}. The removal of the zero order gives a small improvement in the $y$ and $z$ localization precision, and a small deterioration in the $x$ localization precision. Overall, the improvement in the average localization precision is negligible, and this trend is consistent for a range of photon fluxes and $\alpha$ values. 

As previously discussed, spatial filtering of the zero order without blurring the first order can be achieved through the use of a spectral filter. For this reason, the effect of a 10~nm spectral filter is simulated for completeness, with results shown in Table \ref{tab:PSF_ZeroOrder}. This spectral filter removes approximately 79.6\% of the photons in the case of the green fluorescent beads used here, with a corresponding three- to four-fold deterioration in the associated localization precision.

It has been demonstrated that preserving the zero order does not adversely affect the localization precision. Additionally, it has been shown that the necessary use of a spectral filter, as required to spatially filter out the zero order without blurring the PSF, leads to a dramatic deterioration of the localization precision. Finally, as discussed in the main manuscript, filtering out the zero order will deteriorate the space-bandwidth product of the microscope. For these reasons, we opt to preserve the zero order for the measurements undertaken here. 

\begin{table}
    \centering
    \caption{\textbf{Localization precision obtained with and without zero order, as well as with and without spectral filtering.} $\alpha = $ weighting of twin Airy ePSF; $\frac{P_1}{P_0}$ = Fraction of power in first order; $\#$ = number of photons before spectral filtering; $\Delta f_{\lambda}$ = width of spectral filter; $\Delta x$ = localization precision in $x$; $\Delta y$ = localization precision in $y$; $\Delta z$ = localization precision in $z$. Simulations performed assuming the ePSF light sheet configuration. }
    \label{tab:PSF_ZeroOrder}
    
    \begin{tabular}{c c c c | c c c}
    $\alpha$ & $\frac{P_1}{P_0}$ & $\#$ & $\Delta f_{\lambda}$ & $\Delta x$ (nm) & $\Delta y$ (nm) & $\Delta z$ (nm) \\
    \hline
    5.00 & 0.82 & 50,000 & No & 2.74 & 3.24 & 9.27 \\
    5.00 & 1.00 & 50,000 & No & 2.72 & 3.30 & 9.25 \\
    5.00 & 1.00 & 50,000 & 10~nm & 10.06 & 11.56 & 33.10 \\
    \hline
    7.50 & 0.82 & 10,000 & No & 12.76 & 16.74 & 51.05 \\
    7.50 & 1.00 & 10,000 & No & 12.60 & 16.90  & 50.17  \\
    7.50 & 1.00 & 10,000 & 10~nm & 52.06 & 64.89 & 200.20 \\
    \hline
    7.50 & 0.82 & 25,000 & No & 5.43 & 7.14 & 21.92 \\
    7.50 & 1.00 & 25,000 & No & 5.36 & 7.23 & 21.57 \\
    7.50 & 1.00 & 25,000 & 10~nm & 21.14 & 26.39 & 81.53  \\
    \hline
    7.50 & 0.82 & 50,000 & No & 2.94 & 3.88 & 12.01 \\
    7.50 & 1.00 & 50,000 & No & 2.91 & 3.94 & 11.84 \\
    7.50 & 1.00 & 50,000 & 10~nm & 10.82 & 13.53 & 41.91 \\
    \hline
    10.00 & 0.82 & 50,000 & No & 3.56 & 4.92 & 15.23 \\
    10.00 & 1.00 & 50,000 & No & 3.55 & 5.19 & 15.44 \\
    10.00 & 1.00 & 50,000 & 10~nm & 13.06 & 16.99 & 51.62 \\
    \end{tabular}
\end{table}

\subsubsection*{Comparison of twin Airy and tetrapod PSFs for the light sheet fluorescence microscopy geometry}

A phase mask is incorporated into the collection arm of the light sheet microscope to alter its PSF, with the aim of localizing a point emitter within the light sheet volume. The phase mask in the collection arm is optimized to detect emitters within the 4.7~$\mu$m thick Gaussian light sheet projected by the excitation arm. The presence of a zero order due to the use of an SLM is accounted for in these calculations.

\begin{figure}
    \centering
    \includegraphics[width=0.8\textwidth]{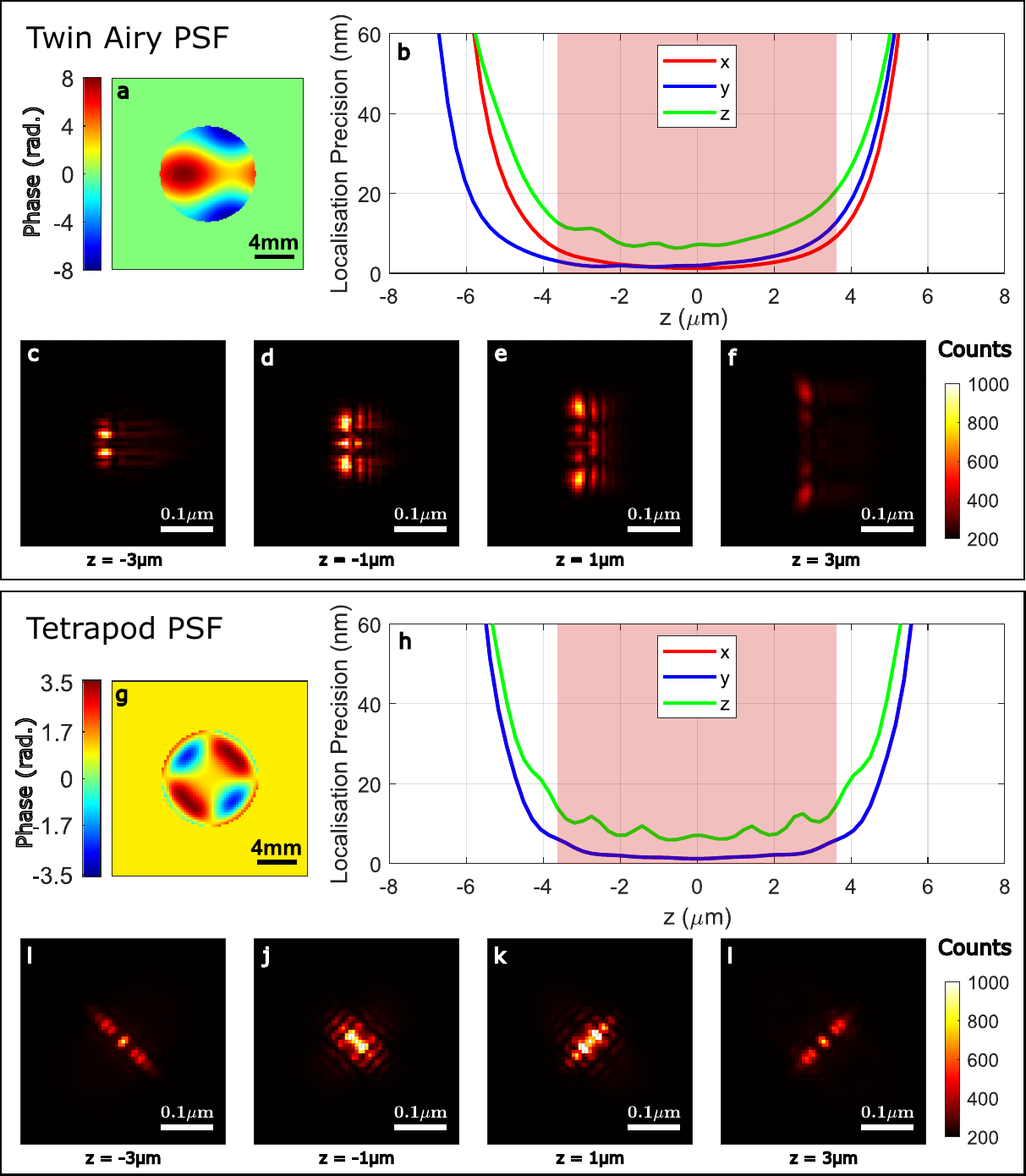}
    \caption{\textbf{Theoretical localization precisions.} The optimized $\alpha=5.35$ Twin Airy mask shown in a) yields the depth-dependent localization precision shown in Fig. b), where the red region corresponds to the light sheet width. Examples of the corresponding PSFs at different axial positions are shown in Figs. c) to f). Similarly, the optimized Tetrapod mask shown in Fig. g) yields the depth-dependent localization precision shown in Fig. h). Examples of the corresponding PSFs at different axial positions are shown in Figs. i) to l). Simulations performed for a photon flux of 50,000 photons.}
    \label{fig:CRLBs}
\end{figure}

Firstly, a twin Airy phase mask \cite{zhou2020twin} is considered, as described by the following equation

\begin{equation}
    \Psi=\alpha \left[ \cos \left(\frac{\pi\cdot u}{R_{pup}} \right)+\frac{1}{2}\sin \left(\frac{\pi\cdot v}{R_{pup}} \right) \right],
    \label{eq:TAphasemask}
\end{equation}

\noindent where $R_{pup}$ is the pupil radius, and $\alpha$ is a scale factor that determines the peak phase modulation. This phase mask approximates two mirrored quasi-cubic phase masks, and so the PSF resembles two Airy beams. The PSF optimisation algorithm indicates that $\alpha=5.35$ is optimal, corresponding to the phase mask shown in Fig. \ref{fig:CRLBs}.

Secondly, a tetrapod phase mask \cite{shechtman2015precise} is considered, consisting of a superposition of Zernike polynomials where weightings are optimized using the PSF optimisation algorithm. The phase mask shown in Fig. \ref{fig:CRLBs} is hence obtained.

A comparison of the localization precisions for the optimized twin Airy mask and the optimized tetrapod mask is provided in Fig. \ref{fig:CRLBs}. The twin Airy mask achieves an average localization precision of 2.71~nm, 3.25~nm and 9.51~nm along the $x$, $y$ and $z$ axes respectively. The tetrapod mask, on the other hand, achieves an average localization precision of 2.22~nm in the $x$ and $y$ directions, and an average localization precision of 8.54~nm in the $z$ direction. Accurate localization is observed for both phase masks within the light sheet region, rapidly increasing outside of the light sheet region. This is as anticipated, as the point emitter is no longer illuminated outside of the light sheet region. The calculations in Fig. \ref{fig:CRLBs} were performed for a photon flux of 50,000 photons incident on the camera. Simulations for lower photon fluxes are provided in Fig. \eqref{fig:localization_DifferentPhotonNumbers} and Table \eqref{tab:localization_DifferentPhotonNumbers}.

The localization precision of the tetrapod mask is only moderately less than the localization precision of the twin Airy mask. However, it is noted that the tetrapod PSF does not provide an intuitive way to infer the $z$ position of the fluorophore. Instead, a maximum likelihood algorithm \cite{shechtman2015precise} or deep learning model needs to be adopted \cite{daly2024high}. For the twin Airy ePSF, on the other hand, the $z$ position of the fluorophore can be directly related to the displacement between the main lobes of the two Airy beams \cite{zhou2020twin}, providing a simple and intuitive approach to inferring axial position. For this reason, the twin Airy ePSF has been used for the remainder of the measurements in this study.

\subsubsection*{Optimized PSFs for different photon fluxes}

\begin{figure}
    \centering
    \includegraphics[width=0.8\textwidth]{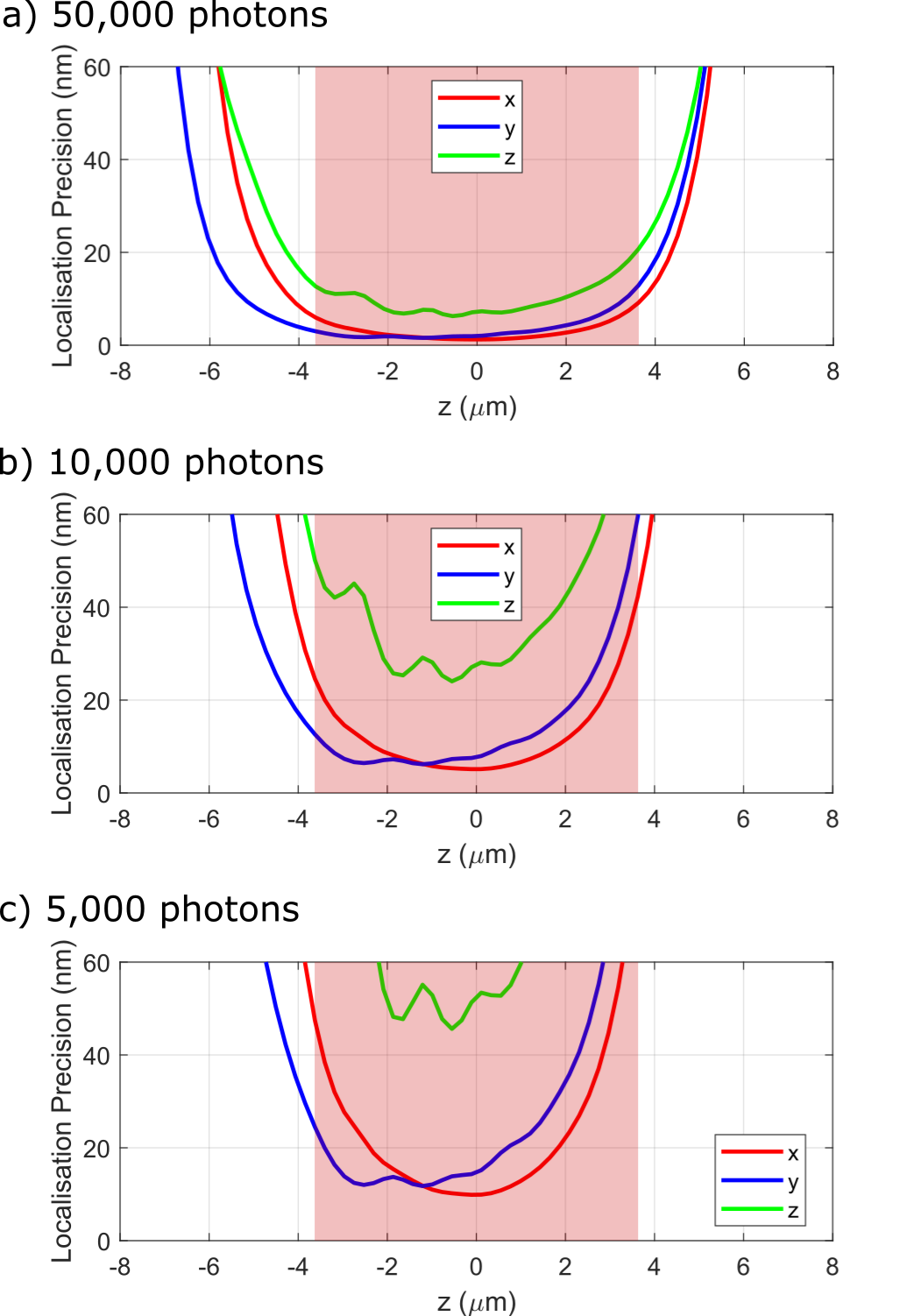}
    \caption{\textbf{Localization precision for lower photon fluxes.} Twin Airy localization precision (see Table \eqref{tab:localization_DifferentPhotonNumbers}) for 50,000, 10,000 and 5,000 photons incident on camera from emitters at the centre of the light sheet shown in a), b) and c) respectively. Figure a) is identical to the results shown in the main manuscript.}
    \label{fig:localization_DifferentPhotonNumbers}
\end{figure}

Twin Airy has been optimized for the ePSF-LSFM geometry. This optimisation was performed for a photon flux of 50,000 photons incident on the camera from photons positioned towards the centre of the light sheet volume. Emitters towards the edge of the light sheet volume see a less intense illumination, and so emit proportionally fewer photons. In Fig. \ref{fig:localization_DifferentPhotonNumbers}, the localization precision of these masks is also evaluated for lower photon fluxes, with minimum and mean localization precisions achieved are reported in Table \ref{tab:localization_DifferentPhotonNumbers}. As anticipated, the localization precision deteriorates as the photon flux is reduced.

\begin{table}
    \centering
    \caption{\textbf{Localization precision achieved for lower photon fluxes for twin Airy ePSF.}}
    \label{tab:localization_DifferentPhotonNumbers}
    \begin{tabular}{c c c c c c c}
    \hline
     & \multicolumn{3}{c}{Min Loc. Precision (nm)} & \multicolumn{3}{c}{Mean Loc. Precision (nm)}\\
    \# Photons & $x$ & $y$ & $z$ & $x$ & $y$ & $z$ \\
    \hline
    50,000 & 1.25 & 1.59 & 6.72 & 2.72 & 3.28 & 9.54 \\
    10,000 & 5.13 & 6.19 & 24.04 & 11.25 & 13.57 & 38.23 \\
    5,000 & 9.88 & 11.76 & 45.62 & 21.72 & 26.21 & 73.30 \\
    \hline

    \end{tabular}
\end{table}

\subsubsection*{Emitter localization in a single plane}

Several steps are required to infer the three-dimensional position of an emitter from its twin Airy ePSF \cite{zhou2020twin}. Briefly:

\begin{enumerate}
    \item twin Airy ePSF main lobes within the image are identified;
    \item the centroid of each detected lobe is determined;
    \item pairs of lobes corresponding to the same emitter are identified;
    \item the $x$ and $y$ coordinates of the emitter are calculated from the average of the lobe coordinates;
    \item the $z$ coordinate of the emitter is calculated from the inter-lobe distance;
\end{enumerate}

When detecting twin Airy main lobes in an image (Step 1), the observed intensity of the lobes depends on the excitation beam intensity. Consequently, emitters positioned towards the centre of a Gaussian beam may yield secondary twin Airy lobes that are brighter than the main lobes of emitters positioned towards the edge of the Gaussian beam. Relatedly, inhomogeneities in biological samples may yield different apparent fluorescent intensities from different spatial positions.

Centroid localization (Step 2) was performed using a standard local-maximum detection routine based on the Crocker–Grier tracking framework \cite{crocker1996methods}, implemented using MATLAB routines adapted from Blair and Dufresne \cite{blair2008matlab}.

When pairing twin Airy lobes (Step 3), it is noted that the characteristic behaviour of the twin Airy ePSF can be leveraged to determine whether two main lobes originate from the same emitter. The centers of the lobes corresponding to the same emitter share the same horizontal coordinate. Additionally, there exists a maximum permissible distance between the main lobes, beyond which they can no longer be attributed to the same emitter \cite{zhou2020twin}. These conditions are particularly useful for distinguishing between the lobes generated by the twin Airy ePSF and the zero order of the SLM.

Inferring the axial position of the emitter from the inter-lobe separation (Step 5) requires a calibration measurement to have been performed. This calibration is performed once for a given microscope configuration using the following procedure:

\begin{enumerate}
    \item identify one or multiple emitters to be used for the calibration procedure;
    \item translate the emitters through the light sheet in known and constant steps;
    \item identify lobes, calculate centroids, pair lobes and calculate inter-lobe distance;
    \item fit calibration curve relating inter-lobe distance to axial position
\end{enumerate}

This structured approach ensures accurate depth retrieval and facilitates high-precision 3D localization of emitters.



\newpage

\begin{longtable}{p{3.2cm} p{10.3cm}}
\caption{\textbf{List of components used in the optical setup.}}
\label{tab:Setupcomponents}\\

\hline
Component & Specifications \\
\hline
\endfirsthead

\hline
Component & Specifications \\
\hline
\endhead

\hline
\endfoot

\hline
\endlastfoot

Laser & Coherent Chameleon Vision S \\

$\lambda$/2 wave plate & Thorlabs AHWP10M-980 --- Ø1" Mounted Quartz Zero-Order Achromatic Half-Wave Plate \\

Polariser crystal & Thorlabs GL10-B --- Mounted Glan-Laser Polarizer, Ø10 mm CA \\

Lens 1 & Thorlabs AC254-050-B-ML --- \(f = 200\) mm, Ø1" Achromatic Doublet \\

Galvo mirror & Thorlabs GVSK1 --- 1D Galvo System Kit, Silver-Coated Mirror \\

Lens 2 & Thorlabs AC254-050-B-ML --- \(f = 200\) mm, Ø1" Achromatic Doublet \\

Mirrors & Thorlabs UM10-AG --- Ø1" Ultrafast-Enhanced Silver Mirror \\

Filter 1 & Thorlabs FELH0725 --- Ø25.0 mm Longpass Filter, Cut-On Wavelength: 725 nm \\

Lens 3 & Thorlabs AC254-200-B-ML --- \(f = 200\) mm, Ø1" Achromatic Doublet \\

Objective 1 & Nikon CFI Plan Fluorite Objective, 10× (0.30 NA, 3.5 mm WD) \\

Chamber & Aluminum \\

Objective 2 & Nikon CFI LWD Plan Fluorite Objective, 16× (0.80 NA, 3.0 mm WD) \\

Lens 4 & Thorlabs AC508-200-B-ML --- \(f = 200\) mm, Ø2" Achromatic Doublet \\

Filter 2 & Semrock 720 nm blocking edge BrightLine\textsuperscript{\textregistered} multiphoton short-pass emission filter \\

Filter 3 & Semrock 694 nm blocking edge BrightLine\textsuperscript{\textregistered} short-pass filter \\

Lens 5 & Thorlabs AC254-100-B-ML --- \(f = 100\) mm, Ø1" Achromatic Doublet \\

Linear polariser & Thorlabs LPVISB100-MP --- Ø25.0 mm Linear Polarizer \\

SLM & Meadowlark E19X12-500-1200 \\

Lens 6 & Thorlabs AC254-200-B-ML --- \(f = 200\) mm, Ø1" Achromatic Doublet \\

Camera & ORCA-Quest 2 qCMOS C15550-22UP \\

\end{longtable}







\end{document}